# Widely tunable, narrow linewidth external-cavity gain chip laser for spectroscopy between 1.0 - 1.1 μm

**Dong K. Shin,**[*] **Bryce M. Henson, Roman I. Khakimov, Jacob A. Ross, Colin J. Dedman, Sean S. Hodgman, Kenneth G. H. Baldwin, and Andrew G. Truscott**

*Laser Physics Centre, Research School of Physics & Engineering, Australian National University, Australian Capital Territory 2601, Australia*
[*]*david.shin@anu.edu.au*

**Abstract:** We have developed and characterised a stable, narrow linewidth external-cavity laser (ECL) tunable over 100 nm around 1080 nm, using a single-angled-facet gain chip. We propose the ECL as a low-cost, high-performance alternative to fibre and diode lasers in this wavelength range and demonstrate its capability through the spectroscopy of metastable helium. Within the coarse tuning range, the wavelength can be continuously tuned over 30 pm (7.8 GHz) without mode-hopping and modulated with bandwidths up to 3 kHz (piezo) and 37(3) kHz (current). The spectral linewidth of the free-running ECL was measured to be 22(2) kHz (Gaussian) and 4.2(3) kHz (Lorentzian) over 22.5 ms, while a long-term frequency stability better than 40(20) kHz over 11 hours was observed when locked to an atomic reference.

## 1. Introduction

Tunable lasers exhibiting narrow spectral linewidths have enabled many technologies that are now ubiquitous in their applications - including optical communications, spectroscopic techniques and frequency metrology. Historically, the field of atomic, molecular and optical (AMO) physics is built on experiments using narrow linewidth lasers, allowing unprecedented control over atoms and molecules. The achievement of laser cooling and trapping of atoms has ultimately led to Bose-Einstein condensation (BEC), creating a degenerate quantum gas at the macroscopic scale [1]. Optical trapping of ions has also been made possible with lasers [2], promising trapped ion quantum computers in the future.

External-cavity lasers (ECL) are the workhorse of many AMO physics labs due to their wide wavelength tunability, narrow spectral linewidth and comparatively low-cost [3]. Their simplicity in design and manufacturability has allowed end-user design and development, especially from laboratories that ultimately utilise the lasers for applications requiring wavelength tunability and low spectral noise [3–11].

Unfortunately, the scarcity in widely tunable sources of high spectral purity at certain wavelengths have hindered research in spectroscopy around such exotic regions. Here we demonstrate the external-cavity laser as a low-cost candidate to one such wavelength range, in particular, the infra-red region around ∼1.0 - 1.1 $\mu$m. As a specific application of our interest, we demonstrate the capability of the laser by performing spectroscopy of metastable helium at 1083 nm with applications to laser cooling and trapping.

Helium has great appeal for atomic and quantum physicists. It is the simplest multi-electron atom, which makes it ideal for tests of quantum electrodynamics such as precision measurement of the tune-out wavelength [12], the fine structure in the $2^3P_J$ manifold [13], and the forbidden $2^3S_1 - 2^1S_0$ transition [14]. In addition, helium in its first excited state ($2^3S_1$) - termed metastable helium (He*) due to its longest atomic excited-state lifetime of ∼8000 s [15] - can be laser cooled to sub-microkelvin temperature for studying the physics of ultra-cold gases and BECs [16, 17]. He* is a favourable atom for many quantum statistical experiments, since the large internal energy of ∼20 eV allows efficient single-atom detection [17].

Laser cooling of He* addresses the $2^3S_1 - 2^3P_2$ atomic transition with a natural linewidth $\Gamma \simeq 1.6$ MHz at 1083.331 nm [17]. The requirement of a narrow linewidth laser at this wavelength in order to achieve BEC of He* has led the ultra-cold He* community to utilise expensive commercial Ytterbium-doped fibre and diode lasers.

The high-cost, limited tuning range and difficulty in end-user modifications/repairs of such systems has motivated us to develop an external-cavity laser for use in ultra-cold He* experiments, as a low-cost, high-performance alternative around 1083 nm.

## 2. Laser design

A variety of ECL designs exist in the literature that trade-off design parameters and architecture to yield the required performance characteristics of the laser (see [19] and [20] for a review of ECLs). For laser cooling and trapping (the intended use of this design), the desired attributes are single spectral and spatial mode operation, fast wavelength modulation to allow high bandwidth frequency stabilisation, narrow spectral linewidth, stable mode-hop free operation and moderate optical power.

Single-angled-facet (SAF) gain chips are favourable as gain media for external-cavity configurations due to their broadband and intrinsically extremely low facet reflectivity, even without an anti-reflection (AR) coating [21]. Thus, a significantly wider wavelength tunability and narrower spectral linewidth have been demonstrated for ECLs using SAF gain chips compared to conventional Fabry-Perot (FP) laser diodes [11, 22, 23], for which AR coating is only effective around the design wavelength. In addition, SAF gain chips are usually constructed with extended active ridge lengths spanning a few millimetres, an order of magnitude longer than FP laser diodes, allowing reductions in injection current induced frequency noise. However, the increased semiconductor device capacitance reduces the modulation bandwidth of the ECL.

Another notable advantage of SAF gain chips in an ECL configuration is the intrinsic output beam pointing stability from the normal facet, enabling a robust, fibre-coupled operation unaffected by adjustments on the Littrow angle. For conventional ECLs utilising FP gain chips for which only the free-space output beam is available, designs such as a fold-mirror set-up must be used to reduce the displacement of the beam [7, 8]. Even then, significant beam steering should still occur for tuning ranges well over ∼10 nm.

We used a commercial, integrated SAF gain module with a built-in fibre-coupled output at the chip's normal facet [18]. The fibre-coupled output simplified the laser housing design for isolation from the environment. This allowed us to construct a hermetically sealed housing and achieve vibration, acoustic and thermal isolation of the laser (Fig. 1(a)). The SAF gain chip had a specified back reflectivity of approximately 10% at the normal facet for output coupling, while the reflectivity at the angled facet was specified to be less than 0.001%. Furthermore, the gain module had a built-in precision thermistor and thermoelectric cooler to allow independent temperature control of the gain chip from the cavity.

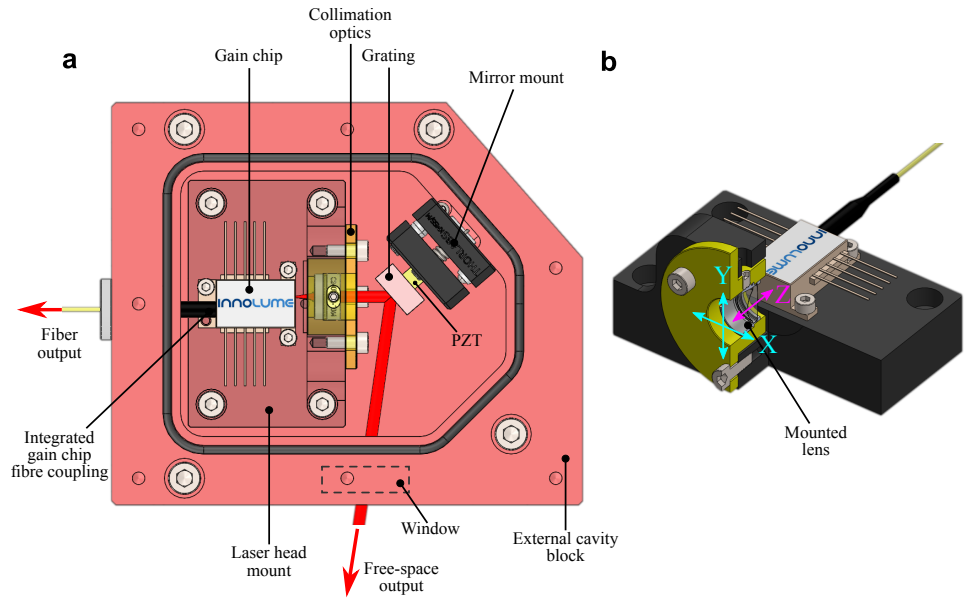

Fig. 1. Mechanical design of the external-cavity laser. (a) The schematic shows the mechanical design of the Littrow configured ECL with a fibre-coupled gain chip. Components not shown in the schematics include the cavity temperature monitoring thermistor, cavity TEC (under the external-cavity block), and the integrated TEC inside the gain chip module. See [18] for a list of components used in the design. (b) An isolated section view of the laser head consisting of the fibre-coupled gain chip and collimation optics. The arrows indicate the degrees-of-freedom in the lens alignment provided by the custom lens mount.

From the many ECL configurations, we adopted the Littrow configuration for its simplicity, as the low number of components helps minimise disturbances from mechanical vibrations, and reduces complexity in alignment. The external-cavity length was designed to be $L_{EC} = 25$ mm for a compact size of the laser and moderate external-cavity free spectral range of $FSR_{EC} = c/2L_{EC} \simeq 6$ GHz.

In order to collimate the free-space output beam from the gain chip, which has a manufacturer specified FWHM divergence angle of 16 deg, a 0.5 NA lens with 8 mm focal length at 1083 nm was used [18]. A customised mount for the collimation lens provided the translational degrees of freedom required to optimise the laser's performance (see Fig. 1(b)). The custom lens mount allowed relatively precise adjustments (~50 $\mu$m) on the distance (Z) between the lens and the gain chip's front facet via the finely pitched thread on the lens tube (see Fig. 1). In contrast, the in-plane (XY) adjustments of the lens was relatively coarse by design and found to be less critical for the alignment (Fig. 1(b)).

A blazed diffraction grating at 1200 lines/mm with 80% efficiency at 1083 nm was selected as the coarse wavelength-selective element to form the external-cavity [18]. To allow coarse tuning of the wavelength and alignment of the back-reflected beam, the grating was fixed onto a kinematic mirror mount [18] with an epoxy adhesive. A piezoelectric chip [18] (PZT) was installed between the grating and the mirror mount to allow high bandwidth and precise wavelength modulation through fine control of the cavity length.

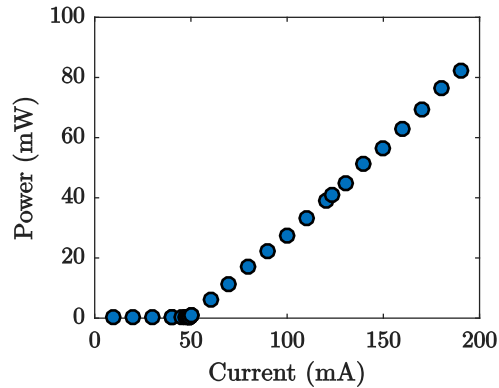

Fig. 2. L-I curve showing the power measured at the fibre-coupled output, versus injection current, at 1083.33(1) nm. The threshold current and slope efficiency were measured to be 50(1) mA and 0.57(1) mW/mA, respectively.

Independent temperature control of the gain chip and the cavity block was achieved from separately placed thermoelectric coolers (TECs) and thermistors [18]: the integrated TEC in the gain module and a TEC module installed under the block. The dual-stage temperature control allowed a fast temperature control of the gain chip independently of the temperature stabilisation of the external-cavity block.

The laser housing - to which the gain chip, collimation optics and kinematic grating body were mounted - was machined from a single block of aluminium, providing a rigid structure that attenuates the environmental noise. The laser was enclosed in an acoustically insulated box built on an optics table and vibrationally isolated from the tabletop via viscoelastic dampers. The box also provided thermal insulation from significant diurnal room temperature fluctuations caused by the laboratory air conditioning system.

## 3. Laser operation

The alignment of the grating and the collimation lens was set to maximise the ECL's output power through the fibre, which was connected to an in-fibre isolator to minimise any back-reflections [18]. A scanning Fabry-Perot interferometer with a free spectral range of ∼1.5 GHz was used to ensure that the laser was operating single-mode during the alignment procedure and the measurements discussed in this paper.

A precise collimation of the beam was critical to achieving the maximum output power. Furthermore, around the optimal location, the ECL's output power was more sensitive to the Z- than the XY-alignment of the lens.

From the measured light output power versus injection current (L-I) curve shown in Figure 2, we determined the threshold current and the slope efficiency to be 50(1) mA and 0.57(1) mW/mA, respectively (at 1083.33(1) nm).

The laser diode current driver used in this experiment limited the maximum injection current to 195 mA, from which the laser's maximum output power was 83(1) mW (Fig. 2). Around 300 mW of output power through the fibre should be achievable by increasing the current to ∼600 mA. We have been operating the laser at a quarter of the gain chip manufacturer's recommended operating current in its continuous operation required for ultra-cold He* experiments, to extend its operational lifetime.

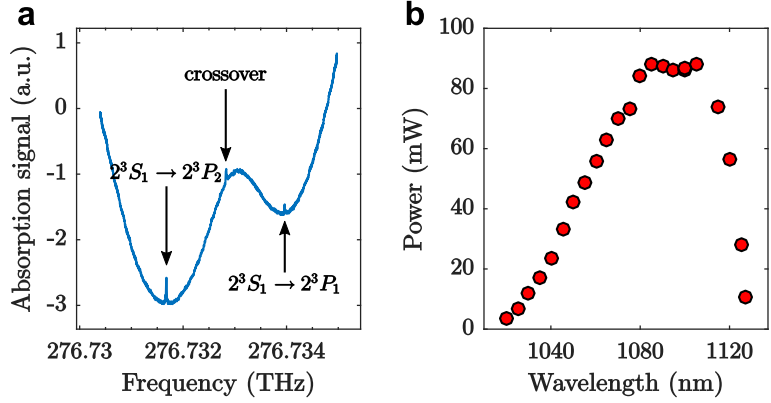

Fig. 3. Wavelength tunability. (a) Saturated absorption spectroscopy of metastable helium obtained from a continuous mode-hop free scan of the laser wavelength over 4.6 GHz (15 pm). The $2^3S_1 - 2^3P_{1,2}$ transitions and the corresponding crossover are indicated in the figure. (b) The plot shows the laser output power ($I_{inj}$ = 195 mA) over the coarse wavelength tuning range. The laser was single-mode at each data point after adjustments. The tuning ranges at FWHM (-3 dB) and FWTM (-10 dB) were determined to be 70(3) nm and 100(3) nm, respectively, with the centre wavelength at 1080(3) nm.

In practice, since the output power of the ECL is insufficient to directly replace our current fibre laser system for the laser cooling of He* operating at roughly 3 W, the ECL serves as the experiment's master laser for seeding a fibre amplifier. Note that a single frequency stabilised ECL could be used to simultaneously seed multiple fibre amplifiers as the seed power requirements are only a few mW.

## 4. Wavelength tunability

The wavelength tunability of an ECL can be characterised either by its mode-hop free (MHF) or coarse tuning ranges. The MHF tuning range refers to the continuous range of wavelengths over which the laser remains single-mode via PZT tuning. In contrast, an adjustment of the grating angle, via the mirror mount, provides a much wider (but discontinuous) coarse wavelength tuning range.

A MHF tuning range of 30(1) pm (7.8 GHz $\simeq 1.3 \cdot FSR_{EC}$) centred around 1083.331(1) nm was observed, limited by the mode competition between longitudinal external-cavity modes when only the cavity length is adjusted [20, 24]. Figure 3(a) shows a saturated absorption spectroscopy (SAS) of the $2^3S_1 - 2^3P_{1,2}$ transitions in helium obtained by scanning the ECL over its MHF tuning range. Synchronous tuning of the grating angle and the PZT displacement is required to further increase the MHF tuning range [24–26].

Figure 3(b) shows the coarse tunability curve obtained from our ECL, demonstrating a FWHM (-3 dB) and FWTM (-10 dB) coarse wavelength tunability of 70(3) nm and 100(3) nm, respectively, around a centre wavelength of 1080(3) nm. The laser operated within 10% of its maximum power when tuned between 1080 nm and 1115 nm, away from which the power decayed asymmetrically, reflecting the amplifier's gain asymmetry. A wider tuning range should be achievable at higher currents due to the broadening of the gain spectrum from populating higher modes in the gain chip's quantum-well.

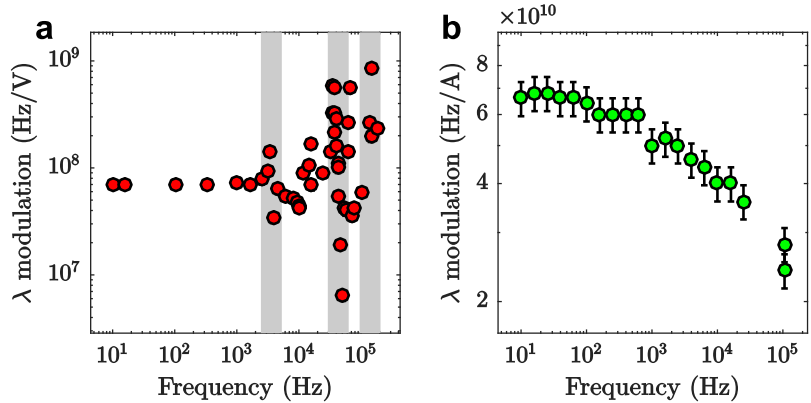

Fig. 4. Laser wavelength modulation response. (a) The frequency response of the laser frequency to modulations of PZT voltage. A flat response of 71(1) kHz/mV was observed below the resonance feature at 3.6 kHz, indicated by the leftmost grey band. The second band located at 40 kHz indicates the predicted mechanical resonance of the grating loaded PZT. The resonance at 140 kHz is most likely an electrical resonance. (b) A similar plot is shown for the modulation of injection current. A flat response of −67(3) kHz/$\mu$A (sign determined from DC adjustments) was observed below 100 Hz and a slow roll-off to the −6 dB point at 37(3) kHz was measured.

## 5. Laser stabilisation

The frequency stability of a laser is an essential requirement for laser cooling and trapping experiments. In addition to improvements in its long-term frequency drifts, a locking scheme with high-bandwidth feedback can provide linewidth narrowing for a free-running laser [24, 27].

In our experiment, we have adopted acousto-optic frequency modulation (see [28] for a detailed description of the scheme) to frequency stabilise the laser to the atomic reference of interest, namely the $2^3S_1 - 2^3P_2$ transition of $^4$He$^*$. In particular, an in-fibre acousto-optic modulator [18] provided the fast dither in laser wavelength at 100 kHz, while an rf-excited helium vapour cell was used as the absolute atomic reference from its Doppler-free absorption features (Fig. 3(a)). A dispersive error signal was then extracted from a phase-locked demodulation of the spectroscopic signal by a lock-in amplifier (LIA) [18]. Frequency stabilisation was achieved from a P-I control of the intra-cavity PZT [18].

The feedback bandwidth of our stabilisation scheme was limited to ∼1 kHz due to the LIA with a maximum reference and demodulation frequency of 100 kHz and 1 kHz, respectively. The laser was locked to the atomic transition and maintained the lock for more than a month. Environmental disturbances including striking the optics table and loud acoustic noises were unable to unlock the laser.

## 6. Wavelength modulation

A number of critical factors influencing the ECL's wavelength were characterised. These include changes in the external-cavity length induced by the PZT and thermal expansion, the laser diode injection current and the temperature of the gain chip. For each parameter, the laser's sensitivity (i.e. the rate of change in lasing frequency, as determined by heterodyne detection, with respect to small changes in the measured parameter) and frequency response were determined and summarised in Table 1.

*6.1. Fast modulation*

The wavelength sensitivity to the PZT voltage measured for our ECL was determined to be 71(3) kHz/mV (Fig. 4(a)). A resonance was observed around 3.6 kHz, most likely from the coupling between the grating loaded piezo and the mirror mount. The usable bandwidth of the PZT was thus limited to the flat-gain region below 3 kHz.

The wavelength modulation transfer function for the injection current is shown in Figure 4(b) and displays no resonance below 100 kHz. The wavelength sensitivity was measured to be −67(3) kHz/$\mu$A, with a −6 dB bandwidth of 37(3) kHz. The measured wavelength sensitivity of the ECL to injection current is around a factor of 50 less than that from conventional FP diode ECLs [3, 29]. The relative insensitivity to current may be due to the large differences in gain stripe lengths between SAF (3 mm) and FP (300 $\mu$m typical) chips, resulting in reduced carrier density changes. The slow response to injection current for our ECL as seen from the low modulation bandwidth, compared to FP diode lasers which can be modulated up to a few gigahertz [3], was most likely due to the larger electrical capacitance from the extended size of the gain chip's active region.

It follows that broadband noise in the injection current should contribute less to the broadening of laser linewidth due to both the relatively insensitive effect on laser wavelength and low modulation cut-off bandwidth. As a consequence, a less stringent noise performance of laser current controllers is required for our ECL to achieve a narrow linewidth compared to conventional ECLs, for which an rms current noise over only a few hundred nA is typically sufficient to broaden linewidths over 100 kHz [8, 30]. For our ECL to achieve a linewidth below 100 kHz, the maximum allowable PZT voltage and injection current noise was calculated to be ∼1.4 mV rms and ∼1.5 $\mu$A rms, integrated over their corresponding bandwidths, respectively.

Table 1. Laser frequency modulation sensitivity. The bandwidth is calculated at the −6 dB gain point. (*Limited by a mechanical resonance)

| Parameter | Sensitivity | Units | Bandwidth |
|---|---|---|---|
| Intra-cavity PZT | 71(3) | kHz/mV | 3* kHz |
| Injection current | −67(3) | kHz/$\mu$A | 37(3) kHz |
| Gain chip temperature | −5.4(4) | MHz/mK | – |
| Cavity temperature | −6.5(5) | MHz/mK | – |

*6.2. Thermal effects*

The sensitivity of the laser wavelength to thermal effects was measured directly with a wavemeter [18]. From incremental temperature changes applied over 200 mK to the gain chip, the laser's wavelength sensitivity with respect to the gain chip temperature was measured to be −5.4(4) MHz/mK.

In order to determine the laser's sensitivity to EC temperature, the drifts in wavelength and cavity temperature were monitored over several hours while all other parameters were stabilised. The measured sensitivity was −6.5(5) MHz/mK, which agrees well with calculations of the cavity thermal expansion.

## 7. Frequency stability

*7.1. Narrow Linewidth*

The spectral linewidth of the ECL was determined from a heterodyne detection of two similarly constructed lasers by a fast photodetector [18]. The resulting beat note was fitted with Gaussian and Lorentzian profiles, from which an individual laser's linewidth can be estimated to lie between FWHM$_{Gaussian}$/$\sqrt{2}$ and FWHM$_{Lorentzian}$/2 [31]. Following Thompson and Scholten [10], each spectrum was divided into the central (1 MHz wide) and tail band for the Gaussian and Lorentzian fits, respectively (see Fig. 5(a) and (b)).

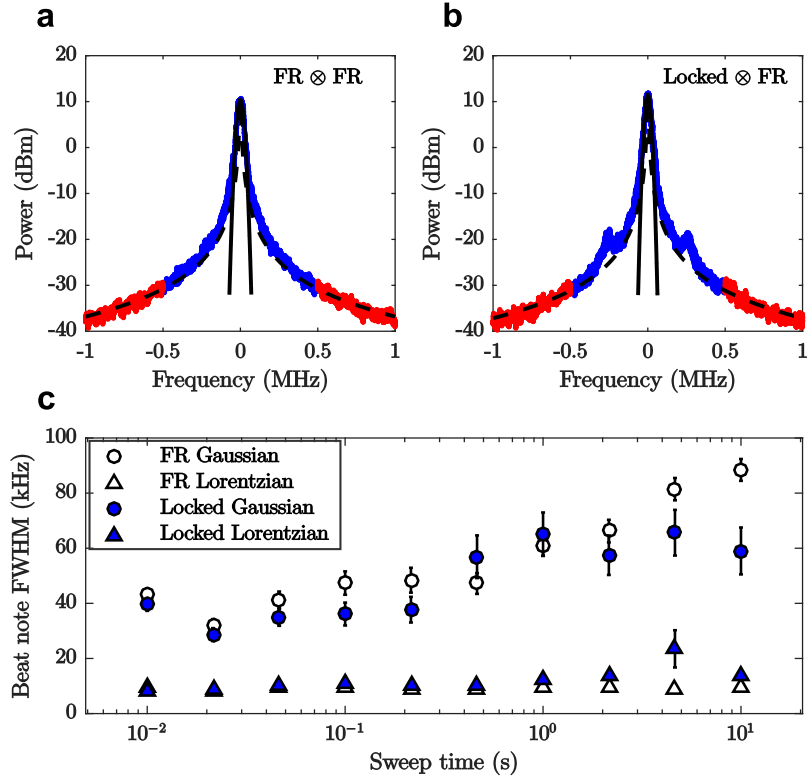

Fig. 5. Heterodyne laser linewidth measurement. (a) Averaged beat note spectrum, taken at 10 ms sweep time, from a heterodyne detection of two identically constructed free-running lasers (FR⊗FR). The beat note is divided into the central 1 MHz region (blue) and remaining tails (red) and separately fitted by a Gaussian (solid curve) and a Lorentzian (dashed curve) profile, respectively. (b) A heterodyne beat note from a free-running and a locked laser (Locked⊗FR) taken over 10 ms sweep time. The sidebands at ~260 kHz were due to the switching noise of the PZT driver used in the locking scheme. (c) The behaviour of heterodyne beat note FWHM when the sweep time of the spectrum analyser is varied between 10 ms to 10 s for the FR⊗FR (unfilled) and Locked⊗FR (filled) cases. The Gaussian peak (circle) was observed to broaden in sweep time in contrast to the Lorentzian tail (triangle) which was almost constant even from three orders of magnitude increase in sweep time. Each data point is an average of fits to 50 individual traces.

Linewidth broadening due to technical noise was minimised for the free-running ECL by positioning the PZT with a low-noise DC voltage supply with a measured noise of ~300 nV rms over 0.2 Hz - 100 kHz bandwidth. In addition, identical low-noise current controllers (custom-built based on the circuitry described in [30]), with noise less than 20 nA rms over the same bandwidth, were used to supply a constant injection current for both lasers. In addition, environmental disturbances to the laser frequency noise were minimised by placing the lasers in an isolated enclosure.

Figure 5(c) shows the beat note FWHM of two free-running lasers and a free-running and a locked laser, at various sweep times (SWTs) of the spectrum analyser [18]. The minimum linewidth for the free-running laser was observed with a fitted Gaussian and Lorentzian linewidths of 22(2) kHz and 4.2(3) kHz, respectively, at 22.5 ms SWT (Fig. 5(c)). No significant changes to the beat note linewidth was seen when one of the ECLs was locked (see Fig. 5(c)). However, small sidebands at 260 kHz were present in the heterodyne spectra with a locked laser (compare Fig. 5(a) and (b)). The sidebands correspond to a narrow-band noise from the PZT driver, required only when locking the laser.

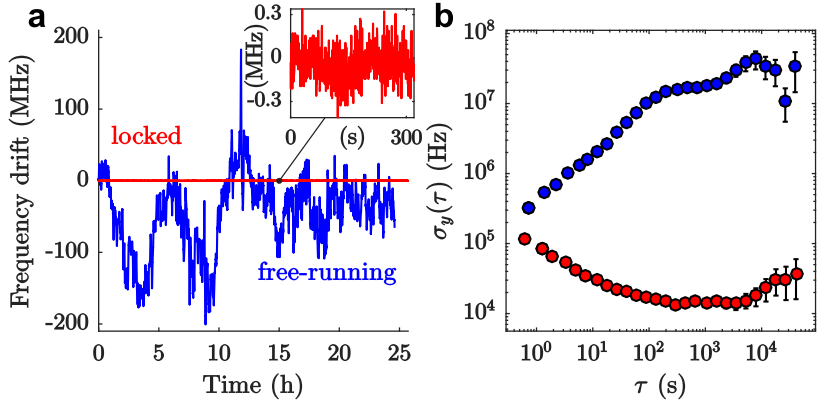

Fig. 6. Long-term laser frequency stability. (a) The frequency drift of the free-running (blue) and the locked (red) laser monitored over a day, sampled at ~1.5 Hz. (b) Allan deviation of the laser frequency. The Allan deviations calculated at an observation time of 11 hours for the free-running and locked laser were 30(20) MHz and 40(20) kHz, respectively.

The free-running laser linewidth increases in sweep time as seen by the trend in the Gaussian fits, due to increasing phase noise contributions to the laser linewidth at longer observation times. In contrast, no such behaviour is seen for the Lorentzian counterpart for which the linewidth was independent of observation time since it is less sensitive to low frequency noise [32]. In addition, the stationary Lorentzian lineshape is consistent with the qualities of intrinsic laser line broadening mechanisms including amplified spontaneous emission of the gain chip and lossy output-coupling of the external-cavity [33].

In order to precisely determine the locked laser linewidth, a delayed self-heterodyne technique [11] or locking of two ECLs is required for the heterodyne detection method discussed here. These steps were not undertaken in our study since the locking scheme did not obviously degrade the spectral linewidth.

We note that an investigation into the laser frequency noise spectrum and noise in the feedback chain is required to better understand the noise sources and their relative contributions to the measured integrated linewidth [34]. In principle, the spectral linewidth of our ECL could be further narrowed from a high-bandwidth feedback scheme, such as locking to polarisation spectroscopy [27], or by improving the current set-up to increase modulation and feedback bandwidth, reduce noise in the feedback chain, and increase the dynamic range of error signal.

### 7.2. Long-term frequency drifts

The long-term frequency drift of the laser was measured by monitoring the beat note frequency with respect to a reference laser (a frequency stabilised fibre laser at 1083.331 nm with ~1 MHz linewidth). The beat note frequency was acquired at ~1.5 Hz from a spectrum analyser and monitored for over 24 hours (Fig. 6). The measured frequency drifts from both a free-running and a locked laser in the time-domain are shown in Figure 6(a).

In order to characterise the frequency drift in the laser, the Allan deviation of the laser frequency drift was calculated from the time-domain measurement [35], allowing the stability of laser frequency to be evaluated over different timescales (Fig. 6(b)). As expected, Figure 6(a) shows that locking the laser significantly improves its long-term frequency stability. The free-running laser exhibited fluctuations of ~100 MHz compared to ~0.1 MHz of the locked laser (see inset of Fig. 6(a)). The fluctuations below ~ 0.5 MHz in the beat note frequency measurement observed at the sampling rate were caused by uncertainty due to the reference laser's broad linewidth. Thus we note that the drift measurement in the locked laser was an overestimate limited by the measurement noise and its true frequency stability will be better.

Figure 6(b) shows the Allan deviation increasing in time for the free-running laser, in contrast to the decreasing trend seen for the locked laser. At timescales longer than 100 s, the frequency stability of the free-running laser was improved by around 3 orders of magnitude by locking to an atomic reference.

The Allan deviation for the free-running and locked laser frequencies was observed to behave asymptotically constant at long timescales. The asymptotes are estimated to be at ~30 MHz and ~20 kHz for the free-running and locked laser, respectively. The asymptotic behaviour in the Allan deviation is expected from the servo control of individual inputs to the laser which should suppress low frequency noise around DC.

The calculated Allan deviation for the frequency drift implies a frequency instability in the free-running laser of 30(20) MHz $\equiv$ 19(13) $\cdot$ $\Gamma$ over 11 hours. Thus, without any frequency stabilisation, the ECL is not suitable for its designed applications. However, the laser's frequency instability was reduced to 40(20) kHz $\equiv$ 0.025(13) $\cdot$ $\Gamma$ over the same time scale, upon locking to an atomic reference. The frequency stability of the locked laser should be no worse than 0.5 $\cdot$ $\Gamma$ for even longer timescales, due to the lock to SAS.

## 8. Conclusions

A high performance external-cavity laser, tunable over 100(3) nm around 1080(3) nm, was constructed and characterised. The laser, based on a single-angled-facet gain chip, demonstrated single-mode operation and a moderate output power of 83(1) mW at 195 mA. In principle, the ECL's tuning range and output power of ~150 nm and ~300 mW should be achievable by increasing the current to ~600 mA.

The free-running ECL demonstrated narrow Gaussian and Lorentzian spectral linewidths of 22(2) kHz and 4.2(3) kHz, respectively. A fast control of the wavelength was provided from the modulation of PZT voltage and injection current with bandwidths of 3 kHz and 37(3) kHz, respectively. A simple laser stabilisation scheme improved the long-term frequency stability of the laser by 3 orders of magnitude to better than 40(20) kHz over 11 hours, while no significant effect was observed on the linewidth.

The total cost of parts for constructing our ECL was under $5000, at least a factor of three cheaper than commercially available lasers of comparable performance. The ECL was relatively insensitive to injection current, hence less stringent noise performance in current controllers is required to achieve narrow linewidth operation. In addition, the ECL was robust and simple to operate, and suitable for laser cooling and trapping of He* experiments as a low-cost, high performance seed laser at 1083 nm. Its wide tunability, while maintaining a narrow linewidth should make it an ideal candidate for spectroscopy in the 1 ~ 1.1 $\mu$m region.

**Funding**

Australian Research Council (ARC) FT100100468, DP120101390, DE150100315, DP160102337

**Acknowledgements**

The authors would like to thank Nicholas Robins for helpful discussions on the laser design and loan of equipment, and Ross Tranter for technical assistance.